\documentclass[aps,pre,twocolumn,10pt,superscriptaddress,footinbib]{revtex4-1}
\usepackage{amsmath,amssymb}
\usepackage{graphicx,color,colortbl}
\usepackage{array,tabularx,booktabs}
\newcolumntype{Y}{>{\centering\arraybackslash}X}

\newcommand{\dif}{\mathrm{d}}%
\newcommand{\fdif}{\operatorname{\delta}}
\newcommand{\Fdif}[2]{\frac{\fdif\!#1}{\fdif\!#2}}
\newcommand{\rs}{\vec{r}\hskip1pt'}%
\newcommand{\rss}{\vec{r}\hskip1pt''}%
\newcommand{\INTcalA}{\int_\mathcal{A}\!\!\!}%
\newcommand{\INTfrakA}{\int_{\mathfrak{A}(\vec r, \phi)}\!\!\!\!\!\!\!\!\!\!\!\!\!}%
\newcommand{\INTfrakC}{\int_{\mathfrak{C}(\vec r, \phi)}\!\!\!\!\!\!\!\!\!\!\!\!\!}%
\newcommand{\INTphi}{\int_0^{2\pi}\!\!\!\!\!\!\!}%

\newcommand{\subsubsubsection}{\paragraph}%
\newcommand{\bulk}{\mathrm{bulk}}
\newcommand{\wall}{\mathrm{wall}}

\begin{document}
\title{Hard rectangles near curved hard walls: tuning the sign of the Tolman length}

\author{Christoph E. Sitta}
\affiliation{Institut f{\"u}r Theoretische Physik II: Weiche Materie,
Heinrich-Heine-Universit{\"a}t D{\"u}sseldorf, D-40225 D{\"u}sseldorf, Germany}

\author{Frank Smallenburg}
\affiliation{Institut f{\"u}r Theoretische Physik II: Weiche Materie,
Heinrich-Heine-Universit{\"a}t D{\"u}sseldorf, D-40225 D{\"u}sseldorf, Germany}

\author{Raphael Wittkowski}
\affiliation{Institut f{\"u}r Theoretische Physik II: Weiche Materie,
Heinrich-Heine-Universit{\"a}t D{\"u}sseldorf, D-40225 D{\"u}sseldorf, Germany}

\author{Hartmut L\"owen}
\affiliation{Institut f{\"u}r Theoretische Physik II: Weiche Materie,
Heinrich-Heine-Universit{\"a}t D{\"u}sseldorf, D-40225 D{\"u}sseldorf, Germany}

\date{\today}

\begin{abstract}
Combining analytic calculations, computer simulations, and classical density functional theory we determine the interfacial tension of orientable two-dimensional hard rectangles near a curved hard wall. Both a circular cavity holding the  particles and a hard circular obstacle surrounded by particles are considered. We focus on moderate bulk densities (corresponding to area fractions up to 50 percent) where the bulk phase is isotropic and vary the aspect ratio of the rectangles and the curvature of the wall. The Tolman length, which gives the leading curvature correction of the interfacial tension, is found to change sign at a finite density, which can be tuned via the aspect ratio of the rectangles. 
\end{abstract}

\maketitle


\section{Introduction}
When a fluid is in contact with a wall, the interfacial tension (also called ``wall tension'') $\gamma$ measures the free-energy cost per boundary area due to the presence of the wall. Many boundary and interfacial effects are governed and controlled by the interfacial tension $\gamma$. For example, the wetting properties of a wall by a liquid droplet in the bulk gas phase depend crucially on the wall-gas, wall-liquid, and bulk liquid-gas interfacial tensions as described by Young's famous equation for the contact angle \cite{Dietrich1988, deGennes2004}.
Moreover, heterogeneous nucleation at the wall is strongly affected by the interfacial tension \cite{SandomirskiALE2011}. Simple classical theory for heterogeneous nucleation \cite{CacciutoAF2004,Sear2007} predicts that the size of the critical nucleus is determined by the degree of undercooling and the interfacial tensions between the wall, the bulk phase, and the nucleating phase \cite{CacciutoAF2003}.

In the simplest case, the wall is planar in three spatial dimensions or a straight line in a two-dimensional system. However, in many practical situations the wall is curved. Examples are provided by spherical obstacles or impurities which can act as a seed for heterogeneous nucleation, by porous materials with a lot of inner curved walls and cavities, and by a rough or patterned substrate \cite{HeniL2000, AllahyarovSEL2015}. This raises the question of the curvature dependence of the interfacial tension $\gamma$. For weak curvature, Tolman suggested the asymptotic series expansion \cite{Tolman1949}
\begin{equation}
\gamma (R) = \gamma (\infty ) \Big(  1 - \frac{ 2\ell_\mathrm{T} }{R} + \mathcal{O}(R^{-2}) \Big)
\label{eq:Tolman}%
\end{equation}
where $R$ is the radius of curvature of the wall, $\gamma (\infty )$ is the interfacial tension for an uncurved wall, and the constant $\ell_\mathrm{T}$, which has the dimensions of a length, is referred to as the {\it Tolman length} \footnote{In the original reference \cite{Tolman1949} Tolman derived the expression $\gamma (R) = \gamma (\infty ) / (  1 + 2\ell_\mathrm{T}/R )$ for a spherical droplet with $|\ell_\mathrm{T}/R| \ll 1$, which corresponds to a fluid in a circular cavity here. Nevertheless, the expansion \eqref{eq:Tolman} is also commonly applied for a fluid surrounding an obstacle \cite{BrykRMD2003, LairdHD2012}.}. Of particular importance is the sign of the Tolman length. If it is negative, there is a free-energy penalty upon bending the wall, whereas a positive Tolman length implies a free-energy decrease for a curved wall. For a flexible wall which can change shape, a positive Tolman length would induce a spontaneous curvature of the wall under appropriate conditions.

Therefore, there is a need to understand the sign of the Tolman length on a microscopic (i.e., particle-resolved) level. This is achieved best for simple model systems of classical statistical mechanics. Hard objects have been studied extensively in this respect as temperature scales out and density is the only relevant thermodynamic parameter \cite{FrenkelS2001,IvlevLMR2012,Dijkstra2014}. In three spatial dimensions, hard spheres near a hard wall have received considerable attention \cite{CourtemanchevS1992,Dijkstra2004}.
The interfacial tension between a planar hard wall and a fluid hard-sphere bulk phase has been explored by computer simulations \cite{HeniL1999,DavidchackL2000,LairdD2007,LairdD2010,BenjaminH2012} and provides an ideal testing ground for the performance of approximations in classical density functional theory (DFT) of inhomogeneous fluids \cite{Evans1979,OhnesorgeLW1994,BraderDE2001,Roth2010,GallardoGAK2012,HaertelOREHL2012}. Subsequent analytic calculations \cite{Urrutia2014}, simulations \cite{LairdHD2012}, and DFT calculations \cite{BrykRMD2003, KoenigRM2004, Blokhuis2013} have considered a curved wall exposed to a hard-sphere fluid and found a negative sign of the Tolman length for hard spheres around a spherical obstacle. Moreover, the Tolman length has been accessed for other interactions such as (modified) Lennard-Jones potentials \cite{BykovZ1999, LeiBYZ2005, StewardE2005, Barrett2006, vanGiessenB2009, BlockDOVB2010, BenjaminH2012, WilhelmsenBR2015} or Yukawa potentials \cite{Barrett1999, BykovZ1999}, at phase boundaries \cite{BlockDOVB2010, TroesterOBVB2012}, and in lattice models \cite{TroesterB2011} \footnote{In some systems with not only excluded volume interactions, such as a Lennard-Jones fluid, the magnitude and sign of the Tolman length are still under debate \cite{ Barrett1999, LeiBYZ2005, Barrett2006, BlokhuisK2006, vanGiessenB2009, BennettB2012, WilhelmsenBR2015}.}.

However, no studies have been done so far for the Tolman length of orientable shape-anisotropic particles, which have a nontrivial rotational degree of freedom.
These particles show more complex structuring near walls as both translational and orientational degrees of freedom are coupled. Although one of the simplest of such systems, namely orientable hard rectangles in two spatial dimensions near a wall, has been intensely studied by means of experiments \cite{GalanisNLH2010,CruzHidalgoZMP2010,AcevedoHZMP2013, MuellerdlHRH2015}, simulations \cite{TriplettF2008, delasHerasV2014, GeigenfeindRSdlH2015, OettelKDESHG2016}, DFT calculations \cite{MartinezRaton2007, Chen2013, GonzalezPintoMRV2013, MuellerdlHRH2015, OettelKDESHG2016}, and other theories \cite{EvertsPSvdSvR2016}, the curvature dependence of the interfacial tension in this system has not yet been explored. Here we close this gap. At moderate aspect ratios, hard rectangles exhibit a stable isotropic phase at densities up to at least 50 percent in area fraction (also called ``packing fraction'') but display significantly more complex ordering at higher densities \cite{BatesF2000}. For various aspect ratios and particle number densities corresponding to a bulk isotropic state, we explore in detail the effects of both a concave and a convex wall, corresponding respectively to a circular cavity holding the rectangles and a hard circular obstacle surrounded by rectangles.

Our results are threefold:
first, we show that this model yields an \textit{analytic expression} for the Tolman length at low densities. This is remarkable as any analytic result is helpful in testing approximative theories and understanding qualitative trends directly. Second, we calculate the Tolman length for a range of densities and aspect ratios in the isotropic phase by Monte Carlo (MC) computer simulations and thermodynamic integration. Interestingly, we find a zero in the Tolman length at finite density. This implies that the Tolman length is \textit{tunable} to a large extent via particle shape and density. Finally, we perform DFT calculations for the Tolman length and discuss their performance by comparing the DFT results with our simulation data. For all investigated aspect ratios, we observe good agreement between MC simulations and DFT calculations up to moderate densities.

The paper is organized as follows:
in Sec.\ \ref{chap:analytic} analytic expressions for the Tolman length in systems with a concave and a convex wall, respectively, are derived. Our MC simulations and DFT calculations are described in Sec.\ \ref{chap:methods}. The results of our analytic and numerical calculations are presented and discussed in Sec.\ \ref{chap:results}. Finally, we conclude in Sec.\ \ref{chap:conclusion}.

\section{\label{chap:analytic}Analytic calculations}
\begin{figure}[ht]
\centering
\includegraphics[width=\linewidth]{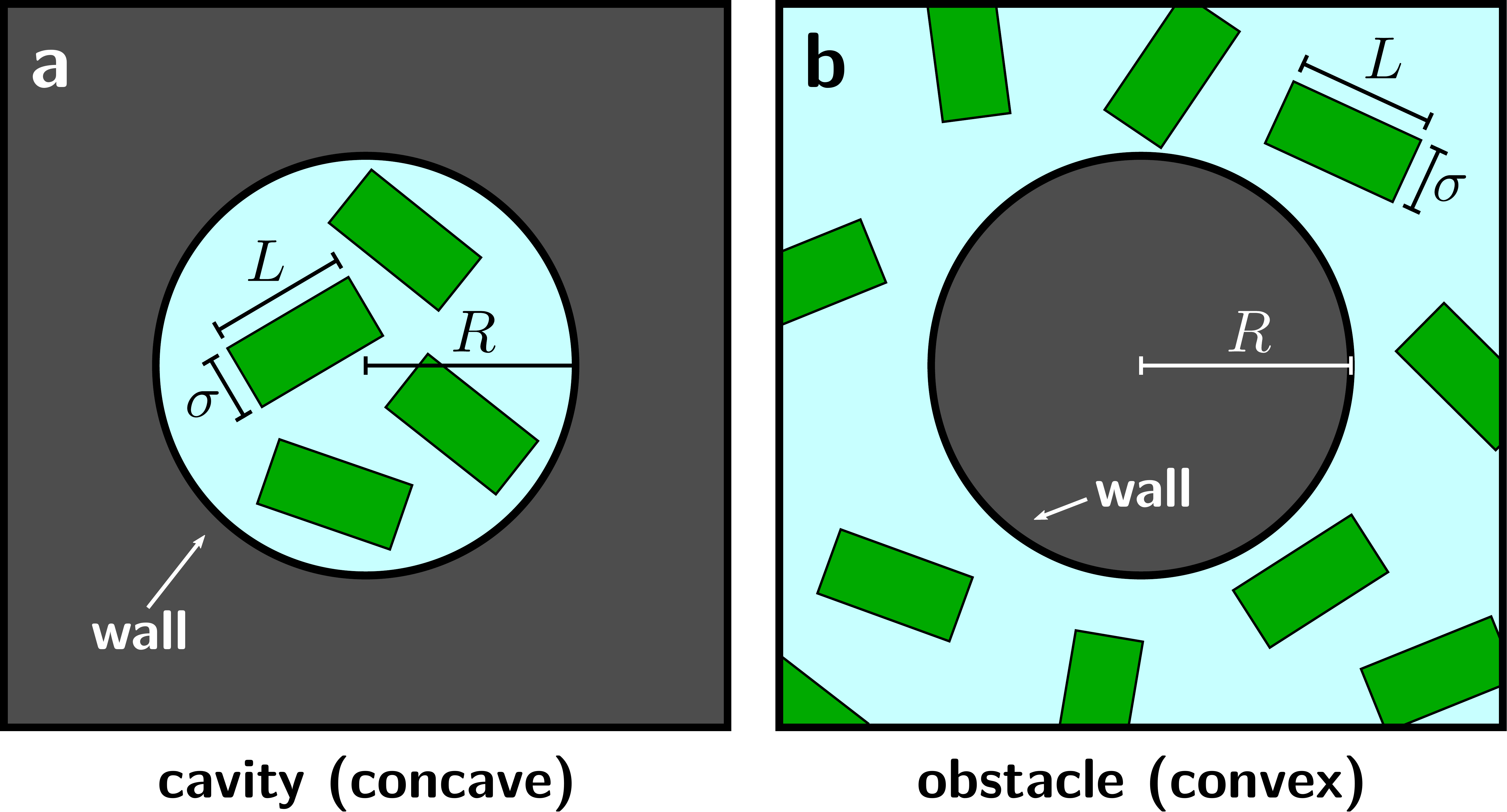}%
\caption{\label{fig:system_sketch}A two-dimensional system of hard rectangular particles with length $L$ and width $\sigma$ either (a) confined by a circular hard wall that forms a cavity with radius $R$ or (b) surrounding a circular hard wall that forms an obstacle with radius $R$.}%
\end{figure}

We study a two-dimensional system of orientable hard rectangular particles with length $L \ge \sigma$ and width $\sigma$ in the presence of a hard unstructured wall. The wall has a constant radius of curvature $R$ so that it forms either a circular cavity (concave wall) containing the rectangular particles (see Fig.\ \ref{fig:system_sketch}a) or a circular obstacle (convex wall) surrounded by the particles (see Fig.\ \ref{fig:system_sketch}b). In the latter case, we assume periodic boundary conditions far away from the circular obstacle. We define the domain $\mathcal{A}$ of the system as the total area accessible to any part of a rectangle (i.e., the light blue areas in Fig.\ \ref{fig:system_sketch}).
The limiting case $R \to \infty$ of an infinite wall curvature radius corresponds to a system with a flat wall, which has already been studied in detail \cite{HeniL1999,LairdD2007,MartinezRaton2007,TriplettF2008, SchoenK2007, LairdD2010,SandomirskiALE2011,BenjaminH2012,AcevedoHZMP2013,GeigenfeindRSdlH2015}.
For the three situations of a flat wall, a cavity containing the particles, and an obstacle surrounded by the particles we are interested in the particle number density $\rho(\vec r, \phi)$, which denotes the probability to find a particle with orientation $\phi$ at center-of-mass position $\vec r = (x,y)$, the interfacial tension $\gamma(R)$, and the Tolman length $\ell_\mathrm{T}$. While for high particle concentrations the quantities $\rho(\vec r, \phi)$, $\gamma(R)$, and $\ell_\mathrm{T}$ are difficult to determine analytically, in the low-density limit interactions between the particles can be neglected and analytic results can be obtained. Therefore, in this section we will focus on low densities. We start with considering the ideal-gas limit where particle-particle interactions are completely negligible. Afterwards we extend our results to higher but still small densities 
on the level of a second-order virial expansion. 

Note that we define the interfacial tension $\gamma$, and therefore the Tolman length $\ell_\mathrm{T}$, in the grand-canonical ensemble, i.e., \cite{BrykRMD2003}
\begin{equation}
\gamma = \frac{\Omega_\mathrm{wall} - \Omega_\mathrm{bulk}}{L_\mathrm{wall}} \,,
\label{eq:interfacial_free_energy}%
\end{equation} 
using the grand-canonical free energy of the system in the presence ($\Omega_\mathrm{wall}$) and absence ($\Omega_\mathrm{bulk}$) of a wall of length $L_\mathrm{wall}$, at fixed temperature $T$ and chemical potential $\mu$. Similar definitions can be written down in other ensembles (using, e.g., the Helmholtz free energy), which are equivalent in the thermodynamic limit for both flat walls and circular obstacles. However, in the case of a circular cavity, the length and curvature of the wall are inherently linked to the system size, which leads to an ensemble-dependence of the apparent Tolman length if Eq.\ \eqref{eq:Tolman} is followed directly.

\subsection{\label{chap:low-density-limit}Tolman length in the ideal-gas limit}
In the ideal-gas limit, where particle-particle interactions can be completely neglected, the particle number density in the grand-canonical ensemble is given by 
\begin{equation}
\rho(\vec r, \phi) = \frac{\rho_0}{2 \pi} e^{-\beta U(\vec r, \phi)} 
\label{eq:id_rho_U}%
\end{equation}
with the constant bulk particle number density $\rho_0=e^{\beta\mu}/\Lambda^2$. Here, $\beta = 1/(k_\mathrm{B} T)$ is the inverse thermal energy with Boltzmann's constant $k_\mathrm{B}$ and $\Lambda$ is the thermal de Broglie wavelength corresponding to the particles. $U(\vec r, \phi)$ is the wall potential that describes the interaction of a particle with center-of-mass position $\vec r$ and orientation $\phi$ with the hard wall. This potential is $\infty$ if $\vec r \notin \mathcal{A}$ or if the particle and the wall (partially) overlap and 0 otherwise. The wall potential $U(\vec r, \phi)$ and thus the particle number density $\rho(\vec r, \phi)$ can therefore be determined by simple geometrical considerations.
If $\rho(\vec r, \phi)$ is known, one can calculate the interfacial tension $\gamma$ from Eq.\ \eqref{eq:interfacial_free_energy}. 
Since particle-particle interactions can be neglected in the ideal-gas limit, the grand-canonical free energies $\Omega_\mathrm{wall}$ and $\Omega_\mathrm{bulk}$ are here given by the exact analytic expressions \cite{EmmerichEtAl2012}
{\allowdisplaybreaks\begin{align}%
\begin{split}%
\Omega_\mathrm{wall} &= \frac{1}{\beta} \INTcalA \dif^2 r 
\INTphi \dif \phi \, \rho(\vec r, \phi)\Big(\ln(2\pi {\Lambda}^2 \rho(\vec r, \phi)) - 1 \\ 
&\quad+ \beta U(\vec r, \phi) - \beta\mu\Big)\,,
\label{eq:Grand_potential_wall}%
\end{split} \\ %
\begin{split}%
\Omega_\mathrm{bulk} &= -\frac{A e^{\beta \mu}}{\beta \Lambda^2}
\label{eq:Grand_potential_bulk}%
\end{split}%
\end{align}}%
with the domain area $A=|\mathcal{A}|$.

Together with Eq.\ \eqref{eq:id_rho_U}, inserting Eqs.\ \eqref{eq:Grand_potential_wall} and \eqref{eq:Grand_potential_bulk} into Eq.\ \eqref{eq:interfacial_free_energy} leads to the interfacial tension
\begin{equation}
\gamma = -\frac{p}{L_\mathrm{wall}} \bigg( \frac{1}{2 \pi} \INTcalA \dif^2 r \INTphi \dif \phi \, e^{-\beta U(\vec r, \phi)} - A \bigg)
\label{eq:obtaining_gamma}%
\end{equation}
with the bulk pressure
\begin{equation}
p  = \frac{e^{\beta \mu}}{\beta \Lambda^2}\,. 
\label{eq:p}%
\end{equation}
From $\gamma$ the Tolman length $\ell_\mathrm{T}$ is obtained by the expansion \eqref{eq:Tolman}.
In the following, the quantities $U(\vec r, \phi)$, which gives $\rho(\vec r, \phi)$ when inserted into Eq.\ \eqref{eq:id_rho_U}, $\gamma(R)$, and $\ell_\mathrm{T}$ are given both for circular and rectangular particles in systems with a flat wall, a cavity (concave wall), and an obstacle (convex wall).

\subsubsection{Hard disks}
For disk-shaped particles of radius $R_0$, the orientation $\phi$ of the particles is trivial due to their full rotational symmetry.

\subsubsubsection{\label{chap:circular_flat}Flat wall:}
We consider a flat wall at $x=0$ and circular particles with center-of-mass positions at $x>0$. The wall potential is then given by
\begin{equation}
U(\vec r, \phi) = 
\begin{cases}%
\infty\,,& \text{if }x\le R_0\,, \\
0\,,& \text{if }R_0<x\,. 
\end{cases}%
\end{equation}
Using Eq.\ \eqref{eq:obtaining_gamma}, one obtains the interfacial tension
\begin{equation}
\gamma = \gamma(\infty) = p R_0
\label{eq:gamma_inf_sphere}%
\end{equation}
with the bulk pressure $p$ for a circular particle given by Eq.\ \eqref{eq:p}. 

\subsubsubsection{Cavity (concave wall):}
If a circular particle is inside a circular cavity of radius $R$ centered at $\vec r = \vec 0$, its interaction with the wall of length $2 \pi R$ is described by the potential
\begin{equation}
U(\vec r, \phi) = 
\begin{cases}%
0\,,& \text{if }r< R - R_0\,, \\
\infty\,,& \text{if }R - R_0 \le r 
\end{cases}%
\end{equation}
with $r=|\vec r|$ denoting the distance of the particle's center of mass from the center of the cavity. Using Eq.\ \eqref{eq:obtaining_gamma}, one obtains the interfacial tension
\begin{equation}
\gamma(R) = \gamma(\infty) \Big(1  - \frac{R_0}{2} \frac{1}{R}\Big) \,.
\end{equation}
Note that this expression is exact and no higher-order terms appear. The corresponding Tolman length is $\ell_{\mathrm{T}} = R_0/4$.

\subsubsubsection{\label{chap:sphere_conv}Obstacle (convex wall):}
A circular obstacle with radius $R$ centered at $\vec r = \vec 0$ interacts with a circular particle via the potential
\begin{equation}
U(\vec r, \phi) = 
\begin{cases}%
\infty\,,& \text{if }r\le R + R_0\,, \\
0\,,& \text{if }R + R_0 < r
\end{cases}%
\end{equation}
with $r=|\vec r|$ denoting the distance of the particle's center of mass from the center of the obstacle. The integral
\begin{equation}
A_\mathrm{f} = \frac{1}{2 \pi}\INTcalA \dif^2 r \INTphi \dif \phi \, 
e^{-\beta U(\vec r, \phi)}
\label{eq:A_f}%
\end{equation}
in Eq.\ \eqref{eq:obtaining_gamma} is basically the angle-averaged free area that is accessible for a particle's center of mass. Its calculation simplifies significantly when using $A_\mathrm{f} - A = - (A_\mathrm{ov} - A_\mathrm{o})$ and the following expression for the overlap area $A_\mathrm{ov}$ of an arbitrary convex particle with area $A_\mathrm{p}$ and circumference $O_\mathrm{p}$ and an arbitrary convex obstacle with area $A_\mathrm{o}$ and circumference $O_\mathrm{o}$ \cite{Boublik1975}:
\begin{equation}
A_\mathrm{ov} = A_\mathrm{p} + A_\mathrm{o} + \frac{O_\mathrm{p} O_\mathrm{o}}{2 \pi}\,.
\label{eq:Vf_convex}%
\end{equation}
For the special case of a circular obstacle with radius $R$, the interfacial tension for any convex particle reads according to Eq.\ \eqref{eq:obtaining_gamma}
\begin{equation}
\gamma(R) = p\frac{ A_\mathrm{p} + O_\mathrm{p}R } {2 \pi R}\,.
\label{eq:general_gamma_convex}%
\end{equation}
Inserting $A_\mathrm{p}=\pi R_0^2$ and $O_\mathrm{p} = 2 \pi R_0$ into Eq.\ \eqref{eq:general_gamma_convex} and using Eq.\ \eqref{eq:gamma_inf_sphere}, this simplifies to the interfacial tension for circular particles
\begin{equation}
\gamma(R) = \gamma(\infty) \Big(1 + \frac{R_0}{2} \frac{1}{R}\Big)\,.
\end{equation}
The corresponding Tolman length is $\ell_{\mathrm{T}} = -R_0/4$.

\subsubsection{\label{chap:rect_ideal}Hard rectangles}
The calculation of the free area $A_\mathrm{f}$ becomes more complicated for rectangular particles with length $L$ and width $\sigma$ as their orientation $\phi$ must be considered. Due to the discrete rotational symmetry of the rectangles, only angles $\phi \in [0,\pi/2]$ need to be taken into account. In the following, the diameter of the rectangles is denoted as $D = \sqrt{L^2 + \sigma^2}$ and the angle between the long side of a rectangle and its diagonal is denoted as $\alpha = \arctan(\sigma/L)$. 

\subsubsubsection{\label{chap:rect_flat}Flat wall:}
We consider the same situation as in Sec.\ \ref{chap:circular_flat}, but now for rectangular particles. The angle $\phi$ is defined as the angle between the wall, i.e., the $y$ axis, and the long side of the rectangle. Depending on the rectangle's distance to the wall, only certain angles are allowed for $\phi$, i.e., correspond to $U(\vec r, \phi) < \infty$. The rectangle's center of mass at distance $x$ from the wall must not approach the wall closer than $\sigma/2$. For $x > \sigma/2$ all angles between 0 and a threshold angle
\begin{equation} 
\phi_1(x) = \arcsin(2x/D) - \alpha 
\end{equation}
are allowed, at which a rectangle's corner touches the wall. Additionally, for $x > L/2$ the rectangle can be orthogonal to the wall $(\phi = \pi/2)$ and also rotate around this orientation up to another threshold angle 
\begin{equation} 
\phi_2(x) = -\arcsin(2x/D) + \pi - \alpha
\end{equation}
at which the same corner collides with the wall again. For $x\ge D/2$ the particle cannot overlap with the wall. This results in the following wall potential (with $0 \le \phi \le \pi/2$):
\begin{equation}
U(\vec r, \phi) = 
\begin{cases}
0\,,& \text{if }\sigma/2 < x \le L/2 \;\wedge\; \phi \in \left[0, \phi_1(x) \right] \,, \\
0\,,& \text{if }L/2 < x \le D/2 \\
&\quad\wedge\; \phi \in \left[0, \phi_1(x) \right] \cup \left[\phi_2(x), \pi/2 \right] \,,\\
0\,,& \text{if }D/2 < x\,, \\
\infty\,,&  \text{otherwise}\,.
\end{cases}
\end{equation}%
The interfacial tension is then according to Eq.\ \eqref{eq:obtaining_gamma}
\begin{equation}
\gamma = \gamma(\infty) = p \frac{L+\sigma}{\pi}\,,
\label{eq:gamma_rect_flat}%
\end{equation}
with the bulk pressure $p$ for a rectangular particle given by Eq.\ \eqref{eq:p}.

\subsubsubsection{Cavity (concave wall):}
The wall potential for a rectangular particle in a cavity with radius $R$ centered at $\vec r = \vec 0$ can be written as
\begin{equation}
U(\vec r, \phi) = 
\begin{cases}
0\,, &\text{if }r < R - \frac{D}{2}\,, \\
0\,, &\text{if }R - \frac{D}{2} \le r < \sqrt{R^2 -\frac{\sigma^2}{4}} - \frac{L}{2} \\
&\quad\wedge \; \phi \in \left[0, \phi_\mathrm{3}(r) \right] \cup \left[\phi_\mathrm{4}(r), \frac{\pi}{2} \right] \,,\\
0\,,&\text{if }\sqrt{R^2 -\frac{\sigma^2}{4}} - \frac{L}{2} \le r < \sqrt{R^2 -\frac{L^2}{4}} - \frac{\sigma}{2} \\
&\quad \wedge \; \phi \in \left[0, \phi_\mathrm{3}(r)\right] \,,\\
\infty\,, & \text{otherwise}
\end{cases}
\end{equation}
with  $r=|\vec r|$ denoting the distance of the rectangle's center of mass from the center of the cavity and $\phi$ defined as the angle between $\vec r$ and a short side of the rectangle. The contact angles of the rectangle's corner with the wall are in analogy to the previous section
\begin{align}
\phi_\mathrm{3}(r) &= \arccos \! \Big(\frac{r^2+\frac{D^2}{4}-R^2}{Dr}\Big) - \frac{\pi}{2} - \alpha \,,\\
\phi_\mathrm{4}(r) &= \frac{3 \pi}{2} - \alpha - \arccos \! \Big(\frac{r^2+\frac{D^2}{4}-R^2}{Dr} \Big)\,.
\end{align}
In a circular cavity, the accessible area for a rectangle's center of mass is independent of the orientation $\phi$ due to the rotational symmetry of the cavity. This simplifies the integration in Eq.\ \eqref{eq:obtaining_gamma} and the interfacial tension reads
\begin{equation}
\begin{split}%
\gamma(R) = \frac{p}{2 \pi R} \bigg( &\pi R^2  
- L \sigma  + L\sqrt{R^2 - \frac{L^2}{4}} + \sigma \sqrt{R^2 - \frac{\sigma^2}{4}} \\ 
&\!+ 2 R^2 \arctan\!\Big(\frac{\sigma}{\sqrt{4 R^2 - \sigma^2}}\Big) \\
&\!- 2 R^2 \arctan\!\Big(\sqrt{\frac{4 R^2}{L^2} -1}\Big) \bigg) \,.
\end{split}%
\label{eq:gamma_rect_conc}\raisetag{11ex}%
\end{equation}
The series expansion
\begin{equation}
\begin{split}%
\gamma(R) = \gamma(\infty) \bigg( 1 &- \frac{L \sigma}{2(L+\sigma)} 
\frac{1}{R} - \frac{L^3 + \sigma^3}{24(L+\sigma)} \frac{1}{R^2} \\
&- \frac{L^5 + \sigma^5}{640(L+\sigma)} \frac{1}{R^4} + \mathcal{O} (R^{-6}) \bigg)
\end{split}%
\end{equation}
with respect to $1/R$ at $R \to \infty$ results in the Tolman length $\ell_\mathrm{T} = L \sigma/ (4(L+\sigma))$.

\subsubsubsection{\label{chap:rect_convex}Obstacle (convex wall):} 
The wall potential for a rectangular particle outside of a circular obstacle with radius $R$ centered at $\vec r = \vec 0$ is given by
\begin{equation}
U(\vec r, \phi) = 
\begin{cases}
0\,,&\text{if }R + \frac{\sigma}{2} \le r < \sqrt{R^2 + \frac{D^2}{4} + R \sigma} \\
&\quad\wedge\; \phi \in \left[0,  \phi_\mathrm{5}(r)\right] \,,\\
0\,,&\text{if }\sqrt{R^2 + \frac{D^2}{4} + R \sigma} \le r < R + \frac{D}{2} \\
&\quad\wedge\; \phi \in \left[0,  \phi_\mathrm{6}(r) \right]\,,\\
0\,,&\text{if }R+\frac{L}{2} < r \le \sqrt{R^2 + \frac{D^2}{4} + R L} \\
&\quad\wedge\; \phi \in \left[\phi_\mathrm{7}(r),\frac{\pi}{2} \right] \,,\\
0\,, &\text{if }\sqrt{R^2 + \frac{D^2}{4} + R L} < r \le R + \frac{D}{2} \\
&\quad\wedge\; \phi \in \left[\phi_\mathrm{8}(r), \frac{\pi}{2} \right] \,,\\
0\,, &\text{if }R + \frac{D}{2} < r\,, \\
\infty\,, & \text{otherwise}
\end{cases}
\end{equation}
with $r=|\vec r|$ denoting the distance of the rectangle's center of mass from the center of the obstacle, $\phi$ defined as the angle between $\vec r$ and a short side of the rectangle, and the contact angles
\begin{align}
\phi_\mathrm{5}(r) &= \arccos\!\Big(\frac{R+\frac{\sigma}{2}}{r}\Big)\,,\\
\phi_\mathrm{6}(r) &= \arcsin\!\Big(\frac{r^2+\frac{D^2}{4}-R^2}{Dr}\Big) - \alpha\,,\\
\phi_\mathrm{7}(r) &= \arcsin\!\Big(\frac{R+\frac{L}{2}}{r}\Big)\,,\\
\phi_\mathrm{8}(r) &= \arccos\!\Big(\frac{r^2+\frac{D^2}{4}-R^2}{Dr}\Big) - \alpha + \frac{\pi}{2}\,.
\end{align}
Here, the rectangle is not restricted to touch the wall with a corner (corresponding to the contact angles $\phi_\mathrm{6}(r)$ and $\phi_\mathrm{8}(r)$). It can also touch the wall with its edges. Therefore, additional cases appear in the potential, where $\phi = \phi_\mathrm{5}(r)$ corresponds to a collision with a long edge and $\phi=\phi_\mathrm{7}(r)$ corresponds to a collision with a short edge.
The interfacial tension can be calculated analogously to the situation for a circular particle in Sec.\ \ref{chap:sphere_conv} by inserting  $A_\mathrm{p}=L \sigma$ and $O_\mathrm{p} = 2(L + \sigma)$ into Eq.\ \eqref{eq:general_gamma_convex} and is given by
\begin{equation}
\gamma(R) = \gamma(\infty) \Big( 1 +  \frac{L \sigma}{2 (L + \sigma)} \frac{1}{R} \Big)\,.
\label{eq:Tolman_rect_conv}%
\end{equation}
Note that this analytic result is exact and no terms of higher order in $1/R$ appear.
The corresponding Tolman length is $\ell_{\mathrm{T}} = -L \sigma/(4 (L + \sigma))$.

Comparing the Tolman lengths for disks and rectangles derived above, two features are remarkable. First, the only difference between the Tolman lengths for the cavity and the obstacle is the sign. Second, for both a cavity and an obstacle, the magnitude of the Tolman length is related to the particle's area $A_\mathrm{p}$ and circumference $O_\mathrm{p}$ via $|\ell_{\mathrm{T}}| = A_\mathrm{p}/(2 O_\mathrm{p})$. We note here that in ensembles other than the grand-canonical one, neither of these two features are reproduced.

\subsection{\label{chap:virial}Low-density expansion of the Tolman length}
The grand-canonical partition function $\Xi$ is given by
\begin{equation}
\Xi = \sum_{N=0}^{\infty} \frac{e^{\beta \mu N}}{\Lambda^{2N}} Q_N \,.
\label{eq:partition_func}%
\end{equation}
Here, we have defined the $N$-particle partition function $Q_N$ as
\begin{equation}
Q_N = \frac{1}{(2 \pi)^N N! }\INTcalA \dif^{2N} r \INTphi \dif^{N} \phi \, e^{-\beta U({\vec r}^N,{\phi}^N)}
\end{equation}
with the $N$-particle interaction potential $U({\vec r}^N,{\phi}^N)$. In the limit of low chemical potential $\mu$, the first few terms in the sum over $N$ in Eq.\ \eqref{eq:partition_func} dominate. Expanding up to second order in the fugacity $z = \exp(\beta \mu) / \Lambda^2$, we obtain
\begin{equation}
\begin{split}%
\beta \Omega = -\ln(\Xi) &= -\ln (1 + z  Q_1 + z^2  Q_2 + \mathcal{O}(z^3)) \\
                         &= -z  Q_1 + z^2   \left(\frac{Q_1^2}{2} - Q_2\right) + \mathcal{O}(z^3)\,. 
\end{split}%
\end{equation}
Using Eq.\ \eqref{eq:interfacial_free_energy}, we can now write the interfacial tension $\gamma$ as
\begin{equation}
\begin{split}%
\beta L_\wall \gamma &= 
 z  (Q_1^\bulk - Q_1^\wall) +z^2   \big(Q_2^\bulk -Q_2^\wall \\
& \quad  - \frac{1}{2}( (Q_1^\bulk)^2  -(Q_1^\wall)^2) \big) +\mathcal{O}(z^3)\,.
\end{split}%
\label{eq:gammaapprox}%
\end{equation}
Here, $Q_1^\bulk$ and $Q_1^\wall$ are equal to $A$ and $A_\mathrm{f}$, respectively, with $A_\mathrm{f}$ as defined in Eq.\ \eqref{eq:A_f}. Additionally, $Q_2^\bulk = A(A - A_\mathrm{ex}^\bulk)/2$, with $A_\mathrm{ex}^\bulk$ the orientationally averaged excluded area between two particles in the bulk, which is given by Eq.\ \eqref{eq:Vf_convex} as
\begin{equation}
A_\mathrm{ex}^\bulk = 2 L \sigma + 2(L+\sigma)^2 / \pi\,.
\end{equation}
Thus, the only remaining unknown quantity is $Q_2^\wall$, which can be written as
\begin{equation}
Q_2^\wall = \frac{1}{2 (2\pi)^2}\INTcalA \dif^2 r_1 \INTcalA \dif^2 r_2 \INTphi \dif \phi_1 \INTphi \dif \phi_2\, e^{-\beta (U_1 + U_2 + U_{12})}\,,
\end{equation}
where $U_1$ and $U_2$ represent the interactions of particles 1 and 2 with the wall, respectively, and $U_{12}$ is the pair-interaction potential of the particles. Although this integral is too cumbersome to tackle analytically, it can be rewritten as
\begin{equation}
Q_2^\wall = \frac{A_\mathrm{f}}{2} \left\langle\frac{1}{2\pi} \INTcalA  \dif^2 r_2 \INTphi \dif \phi_2\, e^{-\beta (U_2 + U_{12})} \right \rangle_1\,,
\label{eq:Q2wall}
\end{equation}
where $\left \langle \cdot \right\rangle_1$ denotes averaging over all positions $\vec r_1 \in \mathcal{A}$ and orientations $\phi_1$ of particle 1 which do not correspond to a particle-wall interaction. The expression in the average in Eq.\ \eqref{eq:Q2wall} simply represents the free area available to particle 2 for a given choice of $\vec r_1$ and $\phi_1$. Thus, $Q_2^\wall$ can be written as
\begin{equation}
Q_2^\wall = \frac{A_\mathrm{f}}{2} \left( A_\mathrm{f} - \langle A_\mathrm{ex}^\wall \rangle \right), \label{eq:q2wall}
\end{equation}
where $\langle A_\mathrm{ex}^\wall \rangle $ is the orientationally and translationally averaged excluded area between two particles in the given wall geometry. As $\langle A_\mathrm{ex}^\wall \rangle / A_\mathrm{f}$  is simply the probability that two non-interacting particles overlap in the same wall geometry, it can be numerically measured in  simple two-particle MC simulations. 

Combining Eqs.\ \eqref{eq:gammaapprox} and \eqref{eq:q2wall}, we obtain
\begin{equation}
\beta \gamma = \frac{A - A_\mathrm{f}}{L_\wall}z - \frac{A A_\mathrm{ex}^\bulk  - A_\mathrm{f} \langle A_\mathrm{ex}^\wall \rangle}{2 L_\wall}z^2 + \mathcal{O}(z^3)\,.
\end{equation}
Rewriting this expression in terms of the bulk density 
\begin{equation}
\begin{split}%
\rho_0 &= \frac{\langle N\rangle}{A} = \frac{1}{A} \frac{\sum_{N=0}^{\infty} N z^N Q_N}{\sum_{N=0}^{\infty}z^N Q_N} \\
 &=  z + z^2(2 Q_2^\bulk-A^2)/A + \mathcal{O}(z^3) 
\end{split}%
\end{equation}
yields
\begin{equation}
\begin{split}%
\beta \gamma &=  \frac{A - A_\mathrm{f}}{L_\wall}\rho_0 \\ 
             & \quad\, +\frac{(A - A_\mathrm{f}) A_\mathrm{ex}^\bulk + A_\mathrm{f} (\langle A_\mathrm{ex}^\wall \rangle - A_\mathrm{ex}^\bulk)}{2 L_\wall}\rho_0^2 \\ 
             & \quad\, + \mathcal{O}(\rho_0^3)\,.
\end{split}\raisetag{8ex}%
\label{eq:gamma_analytic}%
\end{equation}
Note that the first term here corresponds to the ideal-gas limit considered in Sec.\ \ref{chap:rect_ideal}, as 
\begin{equation}
\frac{A - A_\mathrm{f}}{L_\wall} = \frac{L + \sigma}{\pi}  \left( 1 + \frac{L \sigma}{2(L + \sigma)}\frac{1}{R} \right)
\label{eq:compare_to_ideal}%
\end{equation}
for the obstacle (the corresponding expression for the cavity can be obtained by substituting $R \to -R$ and adding $\mathcal{O}(R^{-2})$ on the right-hand side of Eq.\ \eqref{eq:compare_to_ideal}). We calculate $\langle A_\mathrm{ex}^\wall \rangle $ and the resulting interfacial tensions $\gamma$ for walls with various radii of curvature and for several different aspect ratios. On this basis, we extract from our results the (linear) low-density behavior of the Tolman length.

\section{\label{chap:methods}Numerical methods}
In the following, we define the orientation $\phi$ of a rectangular particle as the angle measured counterclockwise from the $y$ axis to the long axis of the particle (i.e., the particle is parallel to the $y$ axis for $\phi=0$).
To calculate the particle number density $\rho(\vec r, \phi)$ and interfacial tension $\gamma$ at moderate particle densities, where analytic results are no longer possible, we perform MC simulations and numerical calculations based on DFT. The Tolman length $\ell_\mathrm{T}$ is again determined from the wall-curvature dependence of the interfacial tension $\gamma$. In this section, we describe both numerical approaches in detail.

\subsection{Monte Carlo simulations}
We perform MC simulations of perfectly hard rectangular particles in the grand-canonical ensemble and employ thermodynamic integration to obtain the interfacial tensions \cite{FrenkelS2001}. The simulations are performed at constant domain area $A = |\mathcal{A}|$, constant chemical potential $\mu$, and constant temperature $T$ in the presence of flat or curved walls, as well as in the absence of walls. During each simulation, we measure the average number of particles $\left\langle N \right\rangle$ in the simulation box as well as average density profiles  $\rho(\vec r, \phi)$. Overlaps between rectangles are detected using the separating axis theorem (see, e.g., Ref.\ \cite{GottschalkLM1996}). Simulations are run for at least $10^{10}$ MC steps. For simulations where the particles are not completely confined by a wall, the area of the simulation box is chosen such that $A = 2500 \sigma^2$.

To calculate the interfacial tension $\gamma$, we take the derivative of Eq.\ \eqref{eq:interfacial_free_energy} with respect to the chemical potential $\mu$, and obtain
\begin{equation}
L_\mathrm{wall}\frac{\dif \gamma}{\dif \mu} = 
\left\langle N \right\rangle^\bulk_\mu 
- \left\langle N \right\rangle^\wall_\mu \,.
\end{equation}
Here, $\left\langle N \right\rangle^\bulk_\mu$ and $\left\langle N \right\rangle^\wall_\mu$ indicate the average number of particles in a simulation at chemical potential $\mu$ without and with a wall, respectively.
Integrating with respect to $\mu$ from the low-density limit $\mu = -\infty$, we obtain
\begin{equation}
\gamma = \frac{1}{L_\mathrm{wall}} \int_{-\infty}^\mu \!\!\!\!\!\!\! \dif \mu^\prime \, 
(\left\langle N \right\rangle^\bulk_{\mu^\prime} 
- \left\langle N \right\rangle^\wall_{\mu^\prime})\,.
\label{eq:MC_gamma}
\end{equation}
Note that no additional integration constant is required as $\gamma(\mu = -\infty) = 0$. Thus, in order to calculate $\gamma(\mu)$ in each wall geometry, we integrate a fit to the simulation results $\left\langle N \right\rangle^\bulk_\mu - \left\langle N \right\rangle^\wall_\mu$. We make use of our analytic results for the ideal-gas limit (see Sec.\ \ref{chap:low-density-limit}) in order to improve accuracy at low chemical potential. Finally, to convert $\gamma(\mu)$ to a function of the bulk density $\rho_0$, we simply measure
\begin{equation}
\rho_0(\mu) = \frac{\left\langle N \right\rangle^\bulk_\mu}{A}
\end{equation}
in the simulations without walls.

\subsection{Density functional theory}
In addition to MC simulations, we use DFT calculations in order to obtain density profiles $\rho(\vec r, \phi)$ and free energies.
The Helmholtz free energy $\mathcal{F}$ of the system can be written as the sum of an ideal-gas term $\mathcal{F}_{\mathrm{id}}$ and an excess term $\mathcal{F}_{\mathrm{exc}}$:
\begin{equation}
\mathcal{F} = \mathcal{F}_{\mathrm{id}} + \mathcal{F}_{\mathrm{exc}}\,.
\end{equation}
While the free energy for an ideal gas $\mathcal{F}_{\mathrm{id}}$ is analytically known and given by
\begin{equation}
\mathcal{F}_{\mathrm{id}} = k_\mathrm{B} T \INTcalA \dif^2 r \INTphi \dif\phi 
\,\rho(\vec r, \phi) \big(\ln(\Lambda^2 \rho(\vec r, \phi)) - 1 \big)\,,
\end{equation}
the exact excess term  $\mathcal{F}_{\mathrm{exc}} $ is only known in rare cases (e.g., for a hard-rod fluid in one spatial dimension \cite{Percus1976}) and usually needs to be approximated.

An expression for the excess free energy $\mathcal{F}_{\mathrm{exc}} = k_\mathrm{B} T \INTcalA \,\, \dif^2 r \, \Phi_{\mathrm{exc}}(\vec r)$ for hard rectangles in two spatial dimensions was proposed by Mart{\'i}nez-Rat{\'o}n et al.\ \cite{MartinezRatonVM2005}. It is based on an approximation for the rescaled excess free-energy density $\Phi_\mathrm{exc}(\vec r)$. In order to match both the low-density and the high-density limit, they combined the Onsager approximation \cite{Onsager1949} and fundamental-measure theory (FMT) \cite{Rosenfeld1989}. Their expression for $\Phi_\mathrm{exc}(\vec r)$ also recovers results from scaled particle theory in the uniform limit.

In the scope of FMT, weighted densities $n_i(\vec r) $ are defined as the angle-integrated cross-correlations
\begin{equation}
\begin{split}%
n_i(\vec r) &= \INTphi \dif \phi \, [\rho \star \omega^{(i)}](\vec r, \phi) \\
&= \INTphi \dif \phi \INTcalA \dif^2 r' \, \rho(\rs, \phi) \omega^{(i)}(\rs - \vec r, \phi)
\end{split}%
\end{equation}
of the density profile $\rho(\vec r, \phi)$ with the geometric weight functions
{\allowdisplaybreaks\begin{align}%
\omega^{(0)}(\vec r, \phi) &= \frac{1}{4} \delta\Big(\frac{\sigma}{2} 
- |x_{\phi}|\Big) \delta\Big(\frac{L}{2} - |y_{\phi}|\Big)\,,\\
\omega^{(2)}(\vec r, \phi) &=  \Theta\Big(\frac{\sigma}{2} - |x_{\phi}|\Big) \Theta\Big(\frac{L}{2} - |y_{\phi}|\Big)\,.
\end{align}}%
Here, $\delta(x)$ is the Dirac delta function, $\Theta(x)$ is the Heaviside function, 
$x_{\phi}=x\cos(\phi)-y\sin(\phi)$, and $y_{\phi}=x\sin(\phi)+y\cos(\phi)$.
The approximative rescaled excess free-energy density reads \cite{MartinezRatonVM2005}
\begin{equation}
\begin{split}%
\Phi_\mathrm{exc}(\vec r) = &- n_0(\vec r) \ln(1-n_2(\vec r)) - \frac{n_0(\vec r) n_2(\vec r)}{1- n_2(\vec r)} \\ 
& + \frac{1}{2} \INTphi \dif\phi \,\rho(\vec r, \phi)\,[(1-n_2)^{-1} \star \omega^{(0)} ](\vec r, \phi)  \\
&\quad\times\INTcalA \dif^2 r' \INTphi \dif \phi' \rho(\rs, \phi') f(\vec r - \rs, \phi, \phi')\,,
\end{split}%
\end{equation}
where $f(\vec r - \rs, \phi, \phi') $ is the (negative) Mayer function
\begin{equation}
f(\vec{r} - \rs,\phi,\phi') =
\begin{cases} %
1 \;, &  \text{if particles with coordinates }\\
      & (\vec r, \phi) \text{ and } (\rs,\phi') \text{ overlap,}   \\
0 \;, &  \text{otherwise.} 
\end{cases}%
\label{eq:MayerFunction}%
\end{equation}
To obtain the equilibrium density $\rho_\mathrm{eq}(\vec r, \phi)$, we minimize the grand-canonical free-energy functional
\begin{equation}
\Omega[\rho(\vec r, \phi)] = \mathcal{F}[\rho(\vec r, \phi)] 
- \mu \INTcalA \dif^2 r \INTphi \dif \phi \, \rho(\vec r, \phi)
\label{eq:GP_functional}%
\end{equation}
in real space with respect to $\rho(\vec r, \phi)$ using a Picard iteration scheme \cite{Roth2010} in combination with direct inversion in the iterative subspace (DIIS) \cite{Ng1974,Pulay1980,Pulay1982,KovalenkoTNH1999}.

For fixed values of the chemical potential $\mu$, we calculate the equilibrium densities in the bulk, allowing to translate $\mu$ into the corresponding bulk area fraction $\eta$. Then the equilibrium density profiles for flat and curved walls with several different radii of curvature $R$ are calculated. From the equilibrium profiles, we determine the corresponding grand-canonical free energies using Eq.\ \eqref{eq:Grand_potential_wall} in the presence and Eq.\ \eqref{eq:Grand_potential_bulk} in the absence of a wall. On this basis, the interfacial tensions $\gamma(R)$ are calculated using Eq.\ \eqref{eq:interfacial_free_energy}. According to Eq.\ \eqref{eq:Tolman}, the Tolman length is proportional to the slope of $\gamma(1/R)$ in the limit $1/R \to 0$, which can be accessed by a polynomial fit through the data points for $\gamma(1/R)$. By considering various values of $\mu$, we obtain the Tolman length $\ell_\mathrm{T}(\eta)$ as a function of the bulk area fraction $\eta$. This procedure was repeated for the aspect ratios $L/\sigma = 1$, $2$, $3$, and $4$. Further details on the density-functional minimization are given in Appendix \ref{Ap:DFT}.

\section{\label{chap:results}Results}
\begin{figure*}[ht]
\centering
\includegraphics[width=0.709344876\textwidth]{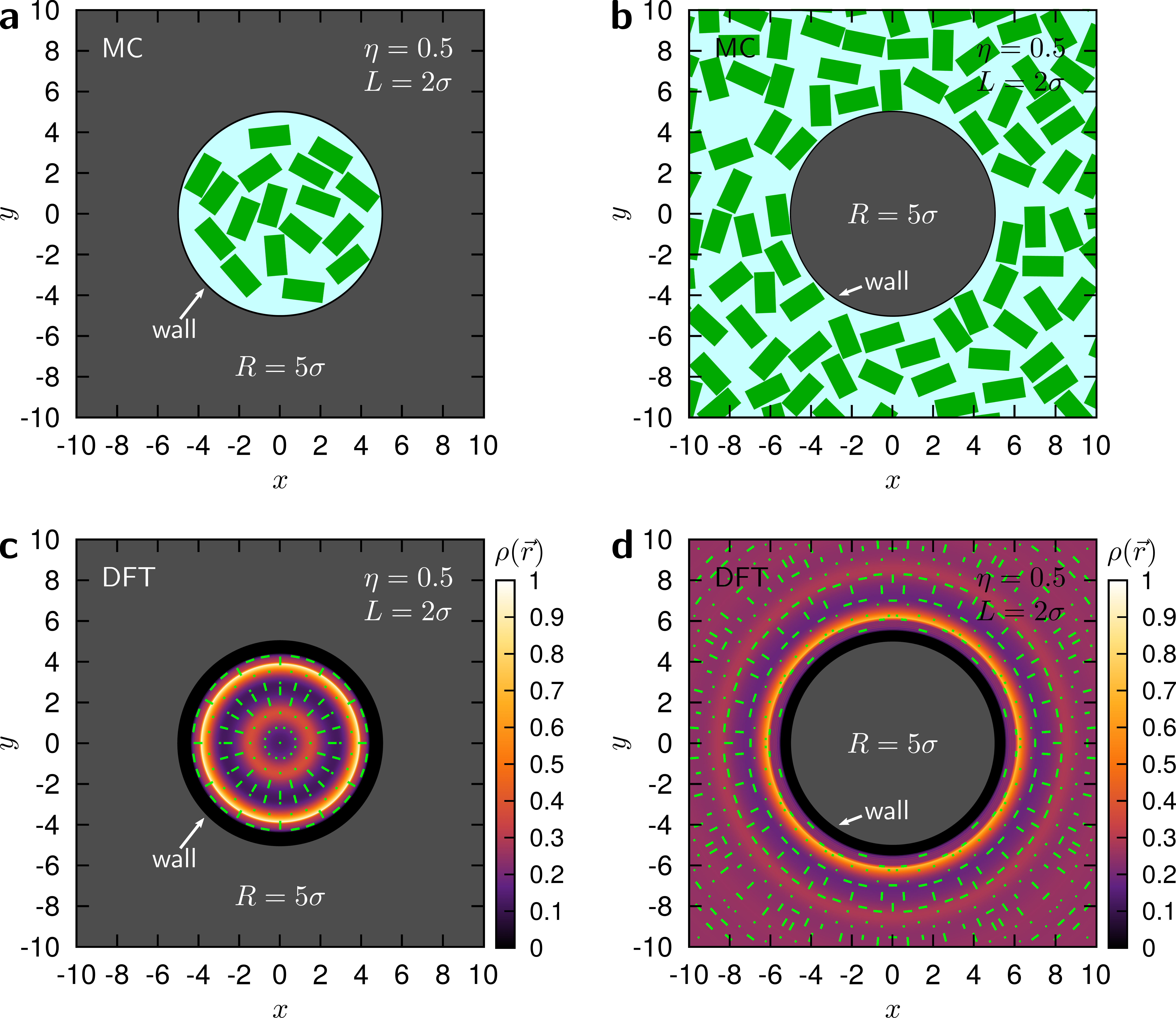}%
\caption{\label{fig:snapshots}Snapshots from MC simulations (top row) and equilibrium density profiles from DFT calculations (bottom row) are shown for rectangular particles with length $L = 2\sigma$ and bulk area fraction $\eta = 0.5$ in a cavity (left column) and around an obstacle (right column) with $R=5\sigma$, respectively. For the DFT results, the orientation-integrated density $\rho(\vec r)$ is shown by the density plots and the green dashes depict the local mean orientation of the particles as well as -- through their length -- the amount of local particle alignment $|S(\vec r)|$. Note that in (b) and (d) the full system is significantly larger than the region shown in these plots.}%
\end{figure*}

Figure \ref{fig:snapshots} shows typical snapshots from our MC simulations (top row) and density profiles from our DFT calculations (bottom row) \footnote{Though only a quarter of the systems was considered in the numerical DFT calculations, we make use of the boundary conditions and show here the full cavity and obstacle for a better illustration.} for equilibrated systems of rectangular particles with $L = 2\sigma$ at bulk area fraction $\eta = 0.5$ in a circular cavity (left column) and around a circular obstacle (right column) with $R=5\sigma$. 
For the DFT results, the orientation-integrated particle number density
\begin{equation}
\rho(\vec r) = \int_0^{2 \pi}\!\!\!\!\!\! \dif\phi \, \rho(\vec r,\phi)
\end{equation}
is shown as a density plot and the orientation field of the particles is depicted with green dashes.
The orientation of the dashes shows the local mean orientation of the particles and the length of the dashes is proportional to the absolute value of the orientational order-parameter field 
\begin{equation}
S(\vec r) =  2 \frac{ \int_0^{2 \pi} \!\! \dif\phi \, \sin^2(\phi-\theta) 
\, \rho(r \hat{u}(\theta),\phi)}{\int_0^{2 \pi} \!\! \dif\phi \, \rho(r \hat{u}(\theta),\phi)} - 1
\end{equation}
with the polar angle $\theta$ and the parametrization $\vec r = r \hat{u}(\theta)$ with $\hat{u}(\theta)=(\cos(\theta),\sin(\theta))$. $S(\vec r)$ describes the amount of local particle alignment relative to the wall with $|S(\vec r)|=1$ for a perfect alignment and $S(\vec r) = 0$ for a uniform distribution of the orientation $\phi$.
In Fig.\ \ref{fig:snapshots}, a layering of the particles near the wall is visible. Like the density field, also the orientation field is rotationally symmetric and shows a damped oscillation as a function of the distance from the wall. 
Near the wall, the local mean orientation of the particles is aligned parallel to the wall. When the distance from the wall is increased, the local mean particle orientation oscillates between an alignment perpendicular ($S(\vec r)<0$) and parallel ($S(\vec r)>0$) to the wall. 

\begin{figure*}
\centering
\includegraphics[width=\textwidth]{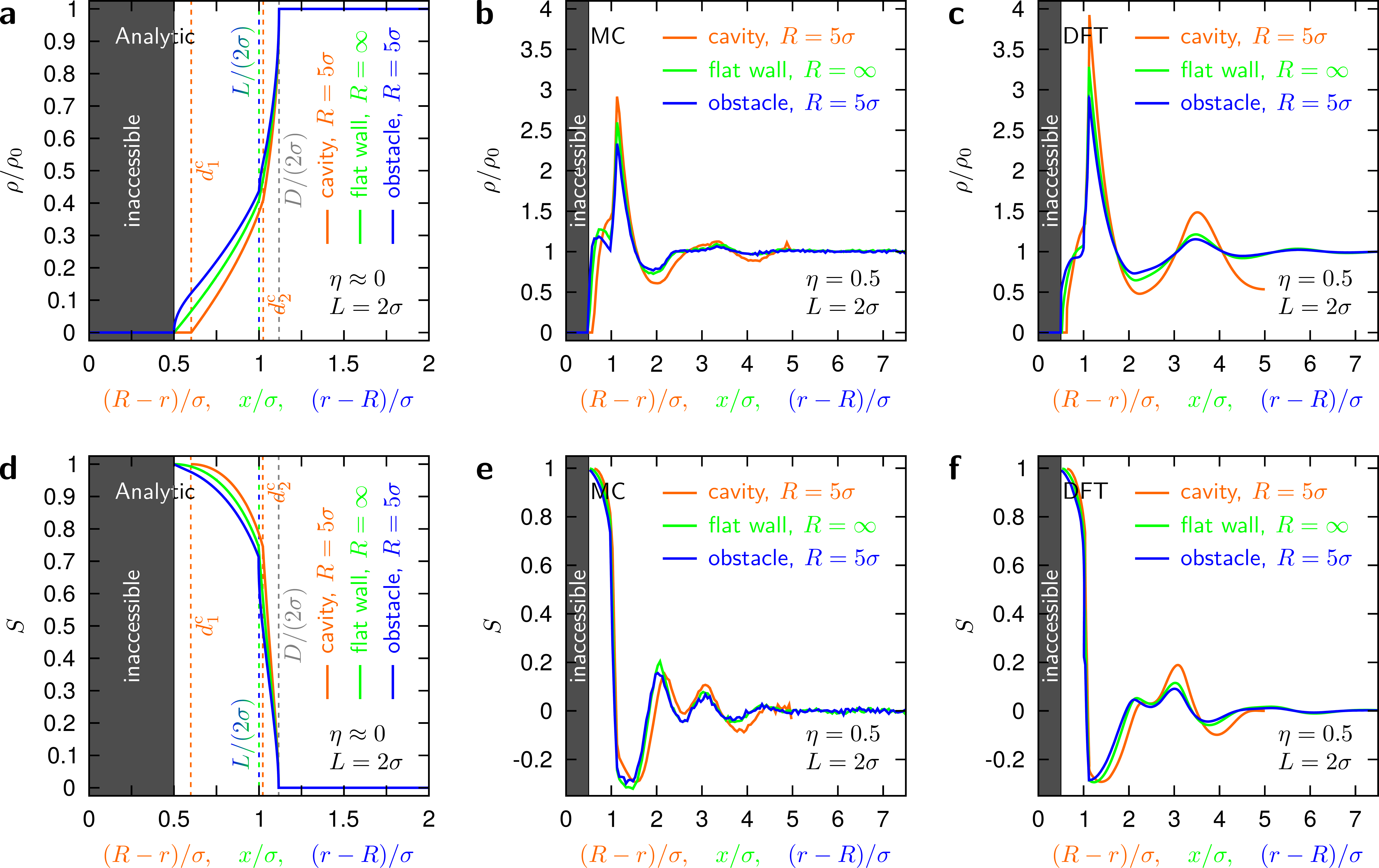}%
\caption{\label{fig:profiles}(a)-(c) The orientation-integrated particle number density $\rho(\vec r)$ and (d)-(f) the orientational order parameter $S(\vec r)$ are shown for rectangular particles with $L=2\sigma$ as a function of the distance from the wall for a circular cavity ($R=5\sigma$, orange), a flat wall ($R=\infty$, green), and a circular obstacle ($R=5\sigma$, blue). 
The columns correspond to (a),(d) analytic results in the ideal-gas limit with $d_1^c = (\sigma/2 + R - \sqrt{R^2 - L^2/4})/ \sigma$ and $d_2^c = (L/2 + R - \sqrt{R^2 - \sigma^2/4})/\sigma$, (b),(e) MC results for $\eta=0.5$, and (c),(f) DFT results for $\eta=0.5$.}
\end{figure*}

In order to compare the different approaches, we calculate from our analytic, MC, and DFT results the orientation-integrated density $\rho(d)$ and the orientational order parameter $S(d)$, where $d$ is the distance of a rectangle's center of mass from the wall in units of $\sigma$, i.e., $d=(R-r)/\sigma$ with $r=|\vec{r}|$ for a cavity, $d=x/\sigma$ for a flat wall, and $d=(r-R)/\sigma$ for an obstacle. These profiles are shown in Fig.\ \ref{fig:profiles} for rectangular particles with $L = 2\sigma$ in the ideal-gas limit (analytic results, left) and at area fraction $\eta = 0.5$ for both MC simulations (middle) and DFT calculations (right). 
Both a cavity (orange) and an obstacle (blue) with $R=5\sigma$ are considered and compared to the limiting case of a flat wall (green). The profiles for the cavity and the obstacle at $\eta = 0.5$ correspond to the snapshots and density profiles shown in Fig.\ \ref{fig:snapshots}.

In the ideal-gas limit (left column in Fig.\ \ref{fig:profiles}), we find differences between the three systems, which can be explained by geometrical considerations. Clearly, the rectangle's center of mass cannot approach a wall closer than half the rectangle's width ($d=0.5$). In the cavity, this inaccessible area is larger, due to the concave curvature of the wall, which prevents the rectangle from touching the wall with its edges. Therefore, this threshold shifts to $d_1^c = (\sigma/2 + R - \sqrt{R^2 - L^2/4})/ \sigma$. At distances smaller than those threshold values, $\rho(d)$ vanishes and $S(d)$ is not defined. As the orientational freedom grows with increasing distance to the wall, $\rho(d)$ increases with $d$ in all three systems. However, we find a qualitatively different behavior for the profiles near a cavity or flat wall, in comparison to that near an obstacle. While $\rho(d)$ is convex in the former two cases, meaning that its second derivative is always positive, we observe a sharp increase of $\rho(d)$ at $d=0.5$ and then a transition from a concave to a convex curve in the latter case. In this concave regime of the density profile around an obstacle, the freedom of rotation of a rectangle is limited by the contact between one of its long edges and the obstacle, rather than its corners, allowing a significantly larger amount of orientational freedom.
When looking at the orientational order parameter very closely to the wall, the rectangles are aligned exclusively parallel to the wall ($S(d)=1$) as only this orientation is possible. Due to the increased orientational freedom further away from the wall, $S(d)$ decreases monotonically with $d$.
As soon as the rectangle's distance from the wall reaches half its length $(d=L/(2\sigma))$ in case of the flat wall or obstacle, or $d_2^c = (L/2 + R - \sqrt{R^2 - \sigma^2/4}) / \sigma$ in case of the cavity, the rectangles may also be aligned orthogonal to the wall and with further increasing distance also rotate around this orientation. This gives rise to a kink in the profiles for both $\rho(d)$ and $S(d)$ in all systems under consideration. For larger distances to the wall, the density profiles increase monotonically with the same qualitative differences between the cavity and the flat wall on the one side, and the obstacle on the other side as observed very close to the wall (see above). This is accompanied with an ongoing monotonic decrease of $S(d)$. As all orientations are allowed for $d \geqslant D/(2 \sigma)$, $\rho(d)$ reaches the bulk density $\rho_0$ and $S(d)$ reaches $0$ at $d=D/(2 \sigma)$. Both $\rho(d)$ and $S(d)$ are constant for $d \geqslant D/(2 \sigma)$. 

We now turn our attention to larger area fractions and focus on $\eta = 0.5$ (middle and right columns in Fig.\ \ref{fig:profiles}). For both MC simulations and DFT calculations, the broadened inaccessible area for the cavity as explained for the ideal-gas limit is retrieved. As in the ideal-gas limit, we find kinks for $\rho(d)$ and $S(d)$ at $d \approx 1$ for both methods \footnote{In the DFT calculations, the orientations $\phi$ are not continuous but discretized, which causes a non-continuous increase in possible orientations and therefore small discontinuities in the profiles for $d<D/(2\sigma)$. As these discontinuities are numerical artifacts with known origin, they were smoothed in Fig.\ \ref{fig:profiles} to show the limit of a continuous $\phi$.}. For larger distances $d$, an accumulation of particles close to the wall as well as a successive layering are clearly visible. Such a layering close to a hard wall is frequently reported in the literature \cite{BrykRMD2003, SchoenK2007, TriplettF2008, MarechalL2013}. Though the amplitudes of the density peaks slightly deviate between MC simulations and DFT calculations, we find very good qualitative agreement when comparing the different wall curvatures, as the relative amplitude differences between the different systems (amplitude for cavity $>$ amplitude for flat wall $>$ amplitude for obstacle) are in agreement. For large distances from the wall, these density fluctuations damp out and the bulk density $\rho_0$ is reached, if enough space is available. We note that in a small cavity (as shown for $R=5\sigma$ in Fig.\ \ref{fig:profiles}c), the bulk density is not reached in the center, which gives rise to strong finite-size effects \footnote{We therefore exclude those small cavities in the calculations of the Tolman length further below.}. In contrast, the bulk reservoir for the flat wall and the obstacle can always be chosen sufficiently large to reach the bulk density in the isotropic phase. In our MC simulations and DFT calculations, we carefully confirmed that the bulk density was reached far away from the wall.

\begin{figure*}[ht]
\centering
\includegraphics[width=0.666267482\textwidth]{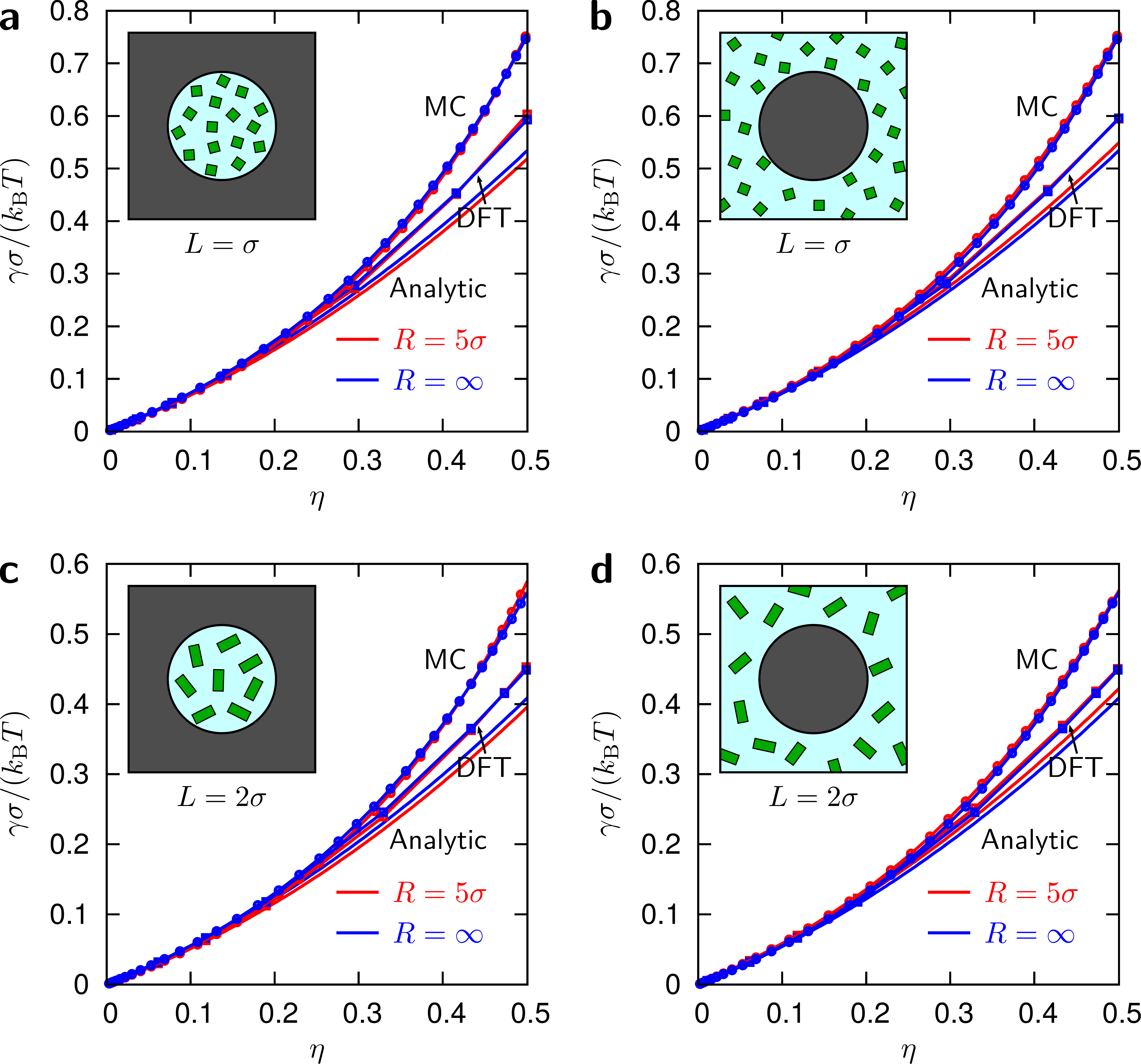}%
\caption{\label{fig:gammas_R}Analytic results from our low-density expansion, MC results, and DFT results for the interfacial tensions $\gamma$ are shown as a function of the bulk area fraction $\eta$ for squares with $L=\sigma$ (top) and rectangles with $L=2\sigma$ (bottom) in a circular cavity (left) and around a circular obstacle (right) with $R=5\sigma$. For small area fractions, the agreement between the results is very good, whereas for larger area fractions, deviations become visible. In the case of the MC and DFT results, even the differences of the curves for $R=5\sigma$ and $R=\infty$ are consistent for both methods.}%
\end{figure*}

Based on our results for the equilibrium density profiles, we determine the interfacial tension $\gamma$ for various aspect ratios $L/\sigma$, bulk area fractions $\eta$, and wall curvatures $\pm 1/R$, using Eq.\ \eqref{eq:gamma_analytic} for our analytic calculations, Eq.\ \eqref{eq:MC_gamma} for our MC simulations, and Eq.\ \eqref{eq:interfacial_free_energy} for our DFT calculations as described in Secs.\ \ref{chap:virial} and \ref{chap:methods}. Figure \ref{fig:gammas_R} shows $\gamma(\eta)$ as obtained by analytic calculations, MC simulations, and DFT calculations for squares with $L=\sigma$ (top row) and rectangles with $L=2\sigma$ (bottom row) in a circular cavity (left column) and around a circular obstacle (right column) with $R=5\sigma$. 
In each plot, we also show the reference case of a flat wall for comparison. We find perfect agreement with our analytic results in the low-density limit. Additionally, at bulk area fractions up to $\eta \simeq 0.3$, we also observe good quantitative agreement between MC and DFT results. For both flat and curved walls, we find a monotonic increase in $\gamma$ with the area fraction. At low densities, a concave curvature of the wall (left column) results in a clear decrease in $\gamma$, whereas a convex curvature (right column) increases $\gamma$, as one would expect from the signs of the Tolman lengths as predicted in the ideal-gas limit (see Sec.\ \ref{chap:low-density-limit}). However, at high densities, the interfacial tension for both the cavity and the obstacle appear to be higher than that for a flat wall. This surprising result occurs for both aspect ratios $L/\sigma = 1$ and $L/\sigma = 2$, and in both the MC simulations and DFT calculations for $R=5\sigma$. We note here that for $R>10\sigma$ the interfacial tension at high densities is lower for the obstacle than for the flat wall. Although this behavior clearly demonstrates that for $R = 5\sigma$ the first-order expansion of $\gamma$ in terms of $1/R$ in Eq.\ \eqref{eq:Tolman} is no longer an accurate approximation, it also strongly suggests that the Tolman length may be strongly dependent on the particle density.

\begin{figure}[ht]
\centering
\includegraphics[width=0.420771274\textwidth]{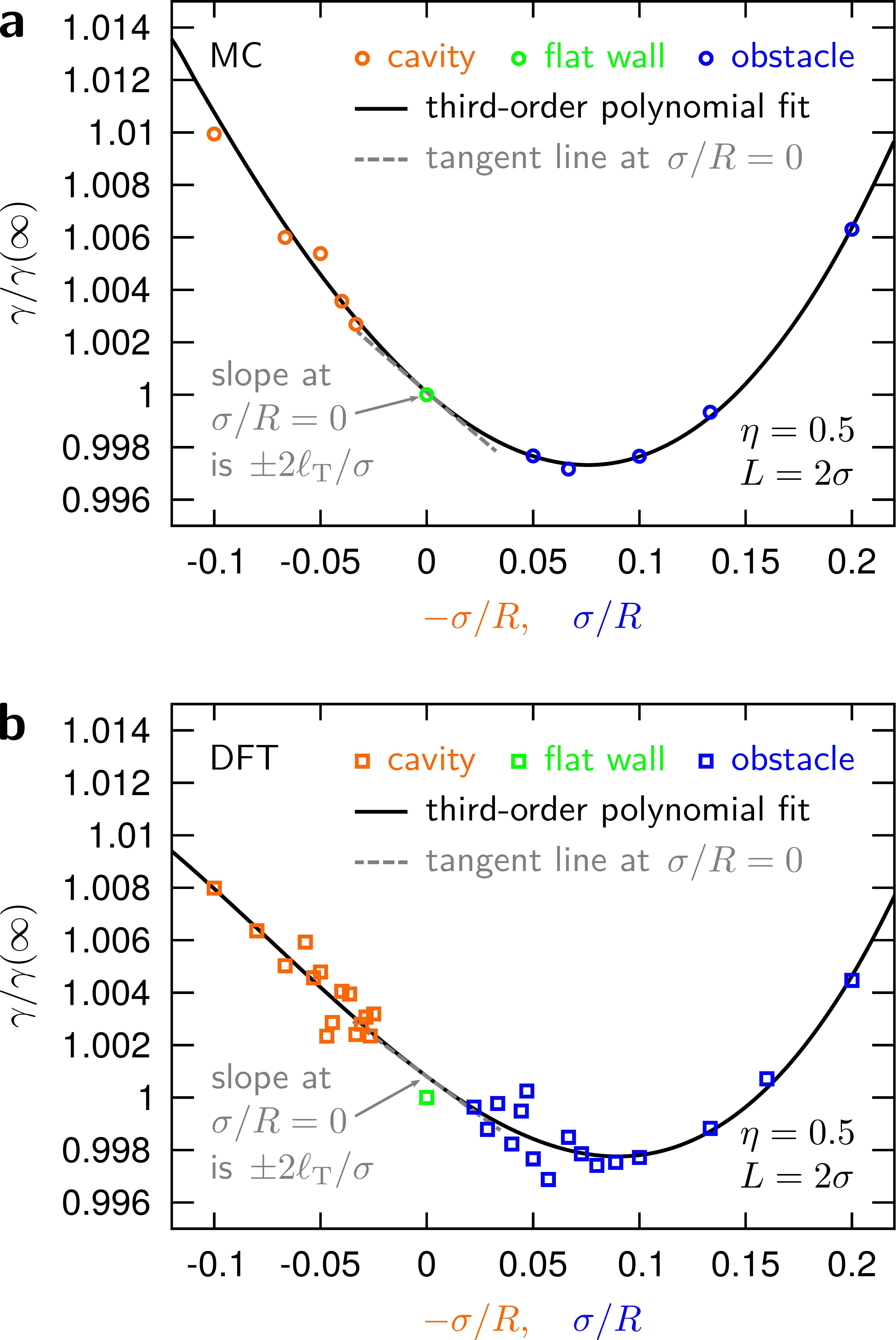}%
\caption{\label{fig:obtaining_tolman}For both (a) MC simulations and (b) DFT calculations, the normalized interfacial tension $\gamma(R)/\gamma(\infty)$ is shown for rectangular particles with $L = 2\sigma$ at bulk area fraction $\eta = 0.5$ in systems with cavities (orange) and obstacles (blue) with various radii of curvature $R$ including a flat wall as limiting case (green, $R=\infty$). Note that the curvature of the cavity is $-1/R$, whereas the curvature of the obstacle is $1/R$. A third-order polynomial fit is also shown. Its slope at $\sigma/R=0$ is $2 \ell_\mathrm{T}/\sigma$ for the cavity and $-2 \ell_\mathrm{T}/\sigma$ for the obstacle, which allows to determine the Tolman length $\ell_\mathrm{T}$.}%
\end{figure}

We therefore now consider the Tolman length in more detail. In both MC simulations and DFT calculations, we obtain the Tolman lengths at finite densities from polynomial fits to the interfacial tension $\gamma(R)$. Consistent with our analytic results, we expected the Tolman lengths for the cavity and the obstacle -- also at higher densities -- to differ only in sign, and not in magnitude. Therefore, we plotted the data for both cavity and obstacle simultaneously, with the cavities corresponding to negative curvatures $-1/R$. Figure \ref{fig:obtaining_tolman} shows an example for $L=2\sigma$ and $\eta = 0.5$. 
In this representation, the Tolman length $\ell_\mathrm{T}$ can be obtained from the slope of $\gamma(R)/\gamma(\infty)$ as a function of the wall curvature $\pm 1/R$, taken in the limit $1/R \to 0$, which we obtained using a single polynomial fit through the data for both the cavities and obstacles to optimize the fit accuracy. In Fig.\ \ref{fig:obtaining_tolman}, $\gamma$ shows a clear negative slope near $1/R = 0$ in both our MC and DFT results, resulting in a negative Tolman length for the cavity and a positive Tolman length for the obstacle \footnote{When using Eq.\ \eqref{eq:Tolman} to relate this slope to the Tolman lengths, one must take into account that in the representation shown in Fig.\ \ref{fig:obtaining_tolman} the interfacial tension $\gamma(R)$ is plotted as a function of the wall curvature, which is $-1/R$ for a cavity and $1/R$ for an obstacle. Therefore, this slope must be multiplied with $\sigma/2$ for the cavity and with $-\sigma/2$ for the obstacle to determine the Tolman length $\ell_\mathrm{T}$}. This is in sharp contrast to the positive Tolman length we find at low densities for the cavity (or negative Tolman length at low densities for the obstacle), and indicates a sign change of the Tolman length as a function of the area fraction for this system. In other words, a bulk particle density exists at which $\ell_\mathrm{T} = 0$, i.e., where the interfacial tension $\gamma(R)$ is, to first order in $1/R$, independent of the radius of curvature $R$. 

\begin{figure}[ht]
\centering
\includegraphics[width=0.47668349\textwidth]{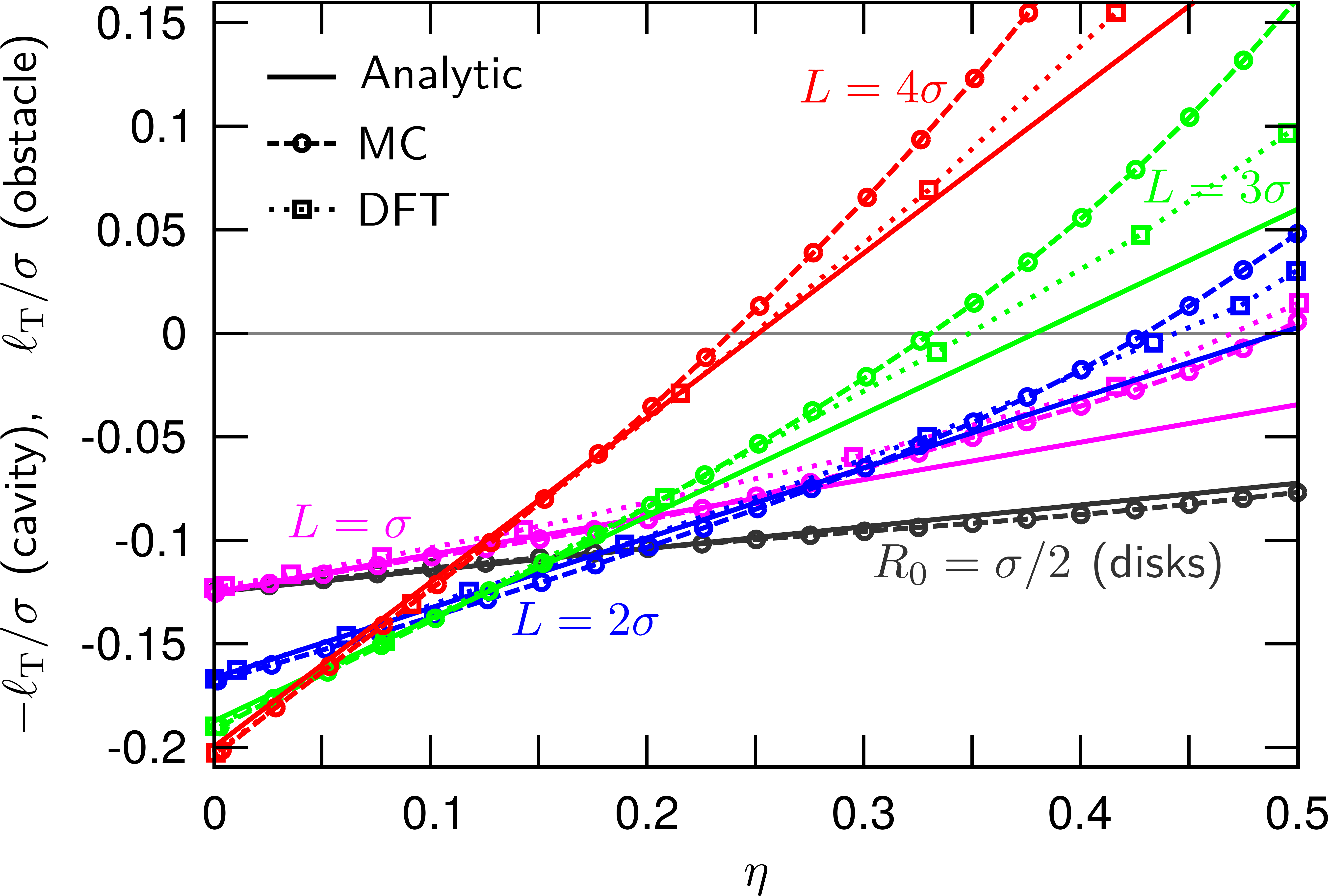}%
\caption{\label{fig:tolman_lengths}Tolman lengths $\ell_\mathrm{T}$ for a fluid inside a circular cavity and for a fluid surrounding a circular obstacle as a function of the bulk area fraction $\eta$. The data are obtained using MC simulations (circles with dashed lines) and DFT calculations (squares with dotted lines) for rectangular particles with aspect ratios $L/\sigma=1$, $2$, $3$, and $4$. In addition, MC results for disks with radii $R_0 = \sigma/2$ are shown. These MC and DFT results are compared with the analytic results from our low-density expansion (solid lines). The agreement is very good for low and intermediate densities. Especially in the ideal-gas limit our analytic results ($\pm 0.125$ for disks and for $L=\sigma$, $\pm1/6 \approx 0.1667$ for $L=2\sigma$, $\pm0.1875$ for $L=3\sigma$, and $\pm0.2$ for $L=4\sigma$) match our numerical results precisely.}%
\end{figure}

In order to examine this intriguing behavior in more detail, we obtained the Tolman lengths for the aspect ratios $L/\sigma=1$, $2$, $3$, and $4$ for various area fractions using MC simulations and DFT calculations and compare these results in Fig.\ \ref{fig:tolman_lengths} with our theoretical results from Sec.\ \ref{chap:analytic}. 
For comparison, we also include MC results for disks of diameter $\sigma$. To maintain readability, we skip distinguishing between cavity and obstacle but instead focus on the obstacle in the following, as the Tolman lengths for the cavity and the obstacle only differ in sign \footnote{The discussion for the Tolman length in the cavity is therefore obtained when exchanging the terms ``positive'' and ``negative'' as well as ``increase'' and ``decrease'', etc.}. At low bulk area fractions $\eta$, we find negative Tolman lengths for all aspect ratios, and observe very good agreement between analytic results, MC simulations, and DFT calculations. With increasing $\eta$, the Tolman length increases monotonically, with higher aspect ratios resulting in a stronger increase. This increase eventually leads to a sign change in $\ell_\mathrm{T}$ for rectangular particles of all investigated aspect ratios. This sign switching is one of the main results of our article and can be observed for lower area fractions as the aspect ratio increases. To our knowledge, a dependence of the sign of the Tolman length on the bulk area fraction was not observed before, and indeed we do not observe this phenomenon for disk-shaped particles \footnote{Although the Tolman length for disks may change sign at area fractions higher than those investigated here, extrapolation would suggest that this does not occur before the onset of the hexatic phase around area fraction $\eta \approx 0.7$ \cite{Mak2006,BernardK2011}.}. This observation for disks in two spatial dimensions is in agreement with previous works, in which no change of sign of the Tolman length was observed for spheres around a cylinder in three spatial dimensions \cite{LairdHD2012}. When scaled accordingly, our results for disks are in qualitative agreement with Fig.\ 2 in Ref.\ \cite{LairdHD2012}. As we do not see a change of sign of the Tolman length for disks, we conclude that the change of sign of the Tolman length is caused by the anisotropy of the particle shape, and not by the restriction to two spatial dimensions.

The effect of particle shape on the density-dependence of the Tolman length, as well as its sign change, are qualitatively captured by the second-order expansion of the interfacial tension in terms of the bulk density in Sec.\ \ref{chap:virial}. This suggests that this behavior can be explained by simple one- and two-particle arguments, even if it occurs at relatively high densities. We recall that up to second order in the fugacity $z$, the surface tension can be written as
\begin{equation}
\begin{split}%
\beta L_\wall \gamma &= (A - A_\mathrm{f})z- \left(A A_\mathrm{ex}^\bulk  - A_\mathrm{f} \langle A_\mathrm{ex}^\wall \rangle \right)\frac{z^2}{2} \\
&\quad\, + \mathcal{O}(z^3) \,. 
\end{split}%
\label{eq:gammaZ}%
\end{equation}
On the right-hand side of this equation, only $A_\mathrm{f}$ and $ \langle A_\mathrm{ex}^\wall \rangle$ depend on the radius of curvature of the wall $R$. The initial negative Tolman length at low density results from the first term of this expansion in $z$. Given the same total available area $A$ and wall length $L_\wall$, the effective free area $A_\mathrm{f}$ is smaller for convex than for flat walls, resulting in a higher interfacial tension $\gamma$. This corresponds to a negative Tolman length.

We now consider the second term on the right-hand side of Eq.\ \eqref{eq:gammaZ}. The term $\langle A_\mathrm{ex}^\wall \rangle$ represents the average area excluded by one particle to another particle within the relevant wall geometry. While far away from the wall, the area excluded by the first particle to the second is simply equal to the bulk value $A_\mathrm{ex}^\bulk$, close to the wall a part of this excluded area is inaccessible to the second particle due to its interaction with the wall, ensuring that $\langle A_\mathrm{ex}^\wall \rangle < A_\mathrm{ex}^\bulk$. Since $A_\mathrm{f}$ is also smaller than $A$ in all cases, the second term on the right-hand side of Eq.\ \eqref{eq:gammaZ} is always negative. Moreover, for convex walls, $A_\mathrm{f}$ is again smaller, and the particles are on average closer to the wall than for a concave wall (see Fig.\ \ref{fig:profiles}), resulting in a smaller $\langle A_\mathrm{ex}^\wall \rangle$ as well. Thus, for convex walls, the $z^2$ term in Eq.\ \eqref{eq:gammaZ} is more strongly negative, resulting in a positive contribution to the Tolman length, which becomes more important at higher fugacity $z$ (i.e., at higher bulk density $\rho_0\propto\eta$). This explains the positive slope of the Tolman length $\ell_\mathrm{T}(\eta)$ as a function of the bulk area fraction $\eta$, which at sufficiently high density leads to a sign change. Finally, we note that for longer particles the effect of the curvature on both $A_\mathrm{f}$ and $\langle A_\mathrm{ex}^\wall \rangle$ is stronger, resulting in a stronger positive slope in $\ell_\mathrm{T}(\eta)$, consistent with our observations in Fig.\ \ref{fig:tolman_lengths}.

\section{\label{chap:conclusion}Conclusions}
In conclusion, we combined analytic calculations, computer simulations, and classical density functional theory to calculate the interfacial tension in a two-dimensional fluid of orientable hard rectangular particles near a curved hard wall. We considered particle densities where the bulk phase of the fluid is isotropic and found that the sign and magnitude of the Tolman length, which characterizes the leading-order curvature contribution to the interfacial tension, varies strongly with the particle shape and density. Specifically, we found a transition from negative to positive Tolman length for a fluid around a circular obstacle (and vice versa for a fluid in a cavity) at a density controlled by the aspect ratio of the rectangles. This sign change does not appear for hard disks in the same geometry. 

Our results are in principle verifiable in experiments with sterically stabilized colloidal \cite{LinCPSWLY2000,GalanisNLH2010,HermesVLVvODvB2011,ZvyagolskayaAB2011,BesselingHKdNDDIvB2015,MuellerdlHRH2015,WalshM2015} or granular \cite{CruzHidalgoZMP2010,HernandezNavarroIMST2012} particles on a two-dimensional substrate. However, it should be noted that the particle number density field near the wall is more direct to obtain than the interfacial tension itself, which requires a thermodynamic integration.

For future studies it would be interesting to generalize our results to various directions: first of all, other bulk phases different from the isotropic fluid such as nematic, smectic, and crystalline phases should be considered. This situation is much more complex and here even the case of a planar hard wall is unexplored. In this case, the interfacial tension will depend also on the relative orientation of the wall with respect to the macroscopic nematic director. Second, other shapes of hard particles should be considered both in two and in three spatial dimensions. These will typically exhibit more complex phase diagrams (see, e.g., Ref.\ \cite{BolhuisF1997}). Concomitantly, new classical density functional theories for shape-anisotropic hard particles based on fundamental measure theory \cite{HansenGoosM2009,HansenGoosM2010,MarechalZL2012,MarechalL2013,WittmannMM2015} should be used to access the Tolman length for bodies of more complex shapes. Some of these were already used for planar hard walls \cite{MarechalL2013} and could be applied to more general systems with curved walls.

\acknowledgments
We thank A.\ Voigt for helpful discussions. Financial support from the Deutsche Forschungsgemeinschaft (Project No.\ LO418/20-1) is acknowledged. 
Moreover, F.\ S.\ gratefully acknowledges funding from the Alexander von Humboldt foundation.

\appendix
\section{\label{Ap:DFT}Numerical details on the density functional minimization}
The area of a rectangle with its center at position $\vec{r}=(x,y)$ and with orientation $\phi$ is denoted as $\mathfrak{A}(\vec{r},\phi)$, whereas the corners of the rectangle $\mathfrak{A}(\vec{r},\phi)$ are denoted as $\mathfrak{C}(\vec{r},\phi)$.
The cross correlations in the calculation of the weighted densities are performed in real space as they can be written as corner- and area integrals 
{\allowdisplaybreaks
\begin{align}%
\begin{split}%
n_{0}(\vec{r}) = \INTphi \dif\phi\!\!\:\!
\INTfrakC \dif^{2}r'
\rho(\rs,\phi) \;, 
\label{eq:n0}%
\end{split}\\
\begin{split}%
n_{2}(\vec{r}) = \INTphi \dif\phi\!\!\:\!
\INTfrakA \dif^{2}r'
\rho(\rs,\phi) \;,
\label{eq:n2}%
\end{split}%
\end{align}}%
using the following notation for the corner- and area integrals of a function $g(\vec r, \phi)$:
{\allowdisplaybreaks
\begin{align}%
\begin{split}%
\INTfrakC \dif^{2}r' \, g(\rs, \phi) 
= \INTcalA \dif^{2}r' \, \omega^{(0)}(\rs-\vec r, \phi) g(\rs, \phi) \,, 
\label{eq:int1}%
\end{split}\\
\begin{split}%
\INTfrakA \dif^{2}r' \, g(\rs, \phi) 
= \INTcalA \dif^{2}r' \, \omega^{(2)}(\rs-\vec r, \phi) g(\rs, \phi) \,.
\label{eq:int2}%
\end{split}%
\end{align}}%
The functional derivative of the excess free-energy functional $\mathcal{F}_\mathrm{exc}[\rho(\vec r, \phi)]$, which is needed for the minimization, is given by
\begin{equation}
\begin{split}%
&\Fdif{\beta\mathcal{F}_{\mathrm{exc}}}{\rho(\vec{r},\phi)} = 
-\INTfrakC\dif^{2}r'\, 
\bigg(\!\ln\!\big(1-n_{2}(\rs)\big) +\frac{n_{2}(\rs)}{1-n_{2}(\rs)}\bigg) \\
&\quad-\INTfrakA\dif^{2}r'\, 
\frac{n_{0}(\rs) n_{2}(\rs)}{(1-n_{2}(\rs))^{2}} 
+m_{1}(\vec{r},\phi)\:\! m_{2}(\vec{r},\phi) \\
&\quad+\INTphi\dif\phi'\!\!\:\!\INTcalA\dif^{2}r'\, 
\rho(\rs,\phi') \:\! m_{2}(\rs,\phi') f(\rs-\vec{r},\phi',\phi) \\ 
&\quad+\frac{1}{2}\INTphi\dif\phi''\!\!\:\!\INTcalA\dif^{2}r''\, 
\rho(\rss,\phi'') \:\! m_{1}(\rss,\phi'') \!\!\:\! \\
&\quad\quad\times\int_{\mathfrak{C}(\rss,\phi'')\cap\mathfrak{A}(\vec{r},\phi)}\!\!\!\!\!\!\!\!\!\!\!\!\!\!\!\!\! 
\!\!\!\!\!\!\!\!\!\!\!\!\!\!\!\!\!\!\dif^{2}r'\, \;\;\;\;\;\;\;\;\;\;\;\;\;
\frac{1}{(1-n_{2}(\rs))^{2}}  
\end{split}%
\label{eq:Funktionalableitung}\raisetag{17.4ex}%
\end{equation}
with the auxiliary functions $m_{1}(\vec{r},\phi)$ and $m_{2}(\vec{r},\phi)$ defined as
{\allowdisplaybreaks
\begin{align}%
\begin{split}%
m_{1}(\vec{r},\phi) = \INTphi\dif\phi'\!\!\:\!\INTcalA\dif^{2}r'\, 
\rho(\rs,\phi') f(\vec{r}-\rs,\phi,\phi') \;,
\label{eq:m1}%
\end{split}\\
\begin{split}%
m_{2}(\vec{r},\phi) = \frac{1}{2}\INTfrakC\dif^{2}r'\, \;
\frac{1}{1-n_{2}(\rs)} \;.
\label{eq:m2}%
\end{split}%
\end{align}}%

The grand-canonical free-energy functional \eqref{eq:GP_functional} is minimized in real space with respect to $\rho(\vec r, \phi)$ using the Picard iteration scheme \cite{Roth2010}
\begin{equation}
\begin{split}%
\rho^{(i+1)}(\vec r, \phi) &= (1-\alpha) \rho^{(i)}(\vec r, \phi) \\
&\quad\, + \alpha \frac{1}{\Lambda^2}\exp{\!\Big(\beta \mu - \frac{\delta \beta \mathcal{F_{\mathrm{exc}}}}{\delta \rho(\vec r, \phi)}\Big)}
\end{split}%
\end{equation}
with the mixing parameter $\alpha \le 0.02$ and $\Lambda$ set to $\sigma$. As in previous works \cite{OettelDBNS2012,HaertelOREHL2012}, this iteration is combined with the DIIS \cite{Ng1974,Pulay1980,Pulay1982,KovalenkoTNH1999} to improve the convergence significantly. The iteration is terminated when the relative change in the grand-canonical free energy during the last 50 iteration steps is less than $10^{-8}$: $|\Omega[\rho^{(n)}(\vec r, \phi)]/\Omega[\rho^{(n-50)}(\vec r, \phi)] - 1| < 10^{-8} $. Reducing the threshold value $10^{-8}$ to $10^{-9}$ does not affect the results. Alternatively, the functional $\Omega[\rho(\vec r, \phi)]$ could also be minimized using dynamical density functional theory \cite{MarconiT1999, MarconiT2000, ArcherE2004, EspanolL2009, WittkowskiL2011}.

The discrete orientations $\phi_i$ of the particles are chosen in equidistant steps of $\Delta \phi = 2 \pi/64$. The orientations are shifted by $\Delta \phi/2$ relative to the orientation of the spatial grid (so that $\phi_1  = \pi/64$) to avoid particle orientations parallel to the grid which might be numerically discriminated or favored. Making use of the particle's symmetries, only $16$ (for $L = \sigma$) or $32$ (for $L > \sigma$) different orientations have to be taken into account.

The step size of the spatial grid is chosen as $\Delta x = \Delta y \approx 0.03 \sigma$. Increasing the resolution of the spatial grid has only a negligible effect on the results. For the corners and edges in the corner- and area integrals a bilinear interpolation is used.

For the bulk system, periodic boundary conditions in the $x$- and $y$-directions are used. In the case of the flat-wall system, the periodic boundaries in the $x$-direction are replaced by hard walls at the borders of the system. It is carefully checked that the bulk density is reached between the walls and that the results do not change upon further increasing the distance between the walls (e.g., a wall distance $L_x = 30\sigma $ is used for $L = 2\sigma$ and $\eta = 0.5$).

For systems with curved walls, we placed the center of the cavity or obstacle at $\vec r = \vec 0$. Making use of the rotational symmetry of our system, we only consider a quarter of the full system ($x\ge0$, $y\ge0$) and ``mirror'' the density profiles at the edges of the system, which significantly speeds up the calculations. When calculating the equilibrium density profiles for an obstacle, we ensure that the domain $\mathcal{A}$ is sufficiently large so that all wall-induced fluctuations are damped out at the edges of the system.

We calculate equilibrium density profiles for various values of the chemical potential $\mu$ and consider at least 26 wall curvatures $\pm 1/R$ for each value of $\mu$. To reduce finite-size effects, we limit the radii of curvature to $R>10\sigma$ for the cavity and $R>5\sigma$ for the obstacle. The interfacial tensions $\gamma$ corresponding to the equilibrium density profiles are normalized by the interfacial tension for a flat wall $\gamma(\infty)$ and plotted as a function of the curvature $\pm 1/R$ as described in Sec.\ \ref{chap:results} (see Fig.\ \ref{fig:obtaining_tolman}). To access the slope of the data at $\sigma/R=0$ and thus to determine the Tolman length, we use a least-squares fit to a polynomial in $\sigma/R$ (up to 3rd order for $L \le 3\sigma$ and up to 4th order for $L = 4\sigma$). We do not force the fit to pass through the data point for the flat wall at $\sigma/R=0$, i.e., the value of the polynomial at $\sigma/R = 0$ is not fixed to 1 but kept as a free fit parameter. In our calculations, this value never differs from 1 for more than $0.005$, demonstrating internal consistency of the data.

\bibliography{refs}

\end{document}